\documentclass[prl,article,nofootinbib,twocolumn,preprintnumbers,groupedaddress,amsmath,amssymb,aps,longbibliography]{revtex4-1}

\usepackage{graphicx}% Include figure files
\usepackage{dcolumn}% Align table columns on decimal point
\usepackage{bm}% bold math
\usepackage{multirow}
\usepackage{comment}
\usepackage{url}
\usepackage{epsfig}
\usepackage{amsfonts}
\usepackage{amsmath}
\usepackage{graphicx}
\usepackage{color}
\usepackage{amssymb,amsmath,hyperref,slashed,color,graphicx,bm,cancel,url}
\usepackage[normalem]{ulem}
\usepackage[T1]{fontenc}
\usepackage{relsize}
\usepackage{gensymb}
\usepackage{fontawesome}
\usepackage[dvipsnames,table]{xcolor}

\hypersetup{colorlinks,citecolor= niceblue,linkcolor= niceblue, urlcolor=niceblue}
\definecolor{niceblue}{rgb}{0.1,0.2,0.6}
\definecolor{kjkblue}{rgb}{0.39, 0.589, 0.6914}

\newcommand{\codeloc}{https://github.com/kjkellyphys/muBHNL}

\begin{document}

\preprint{FERMILAB-PUB-21-277-T}

\title{The MicroBooNE Experiment, the NuMI Absorber, and Heavy Neutral Leptons\\
\textcolor{BlueViolet}{\href{\codeloc}{\faGithub}}\vspace{-0.3cm}}

\author{Kevin J. Kelly and Pedro A.N. Machado}
\affiliation{Theory Department, Fermi National Accelerator Laboratory, Batavia, IL 60510, USA}

\date{\today}% It is always \today, today,

\begin{abstract}
    Motivated by the recent search for a Higgs-Portal Scalar decaying inside the MicroBooNE detector, we demonstrate that the same search can be used to constrain Heavy Neutral Leptons. These are gauge-singlet fermions that interact with the Standard Model by mixing with neutrinos only and could be related to the origin of neutrino masses. By recasting the results of the MicroBooNE Collaboration's analysis, we show that, for a Heavy Neutral Lepton that mixes predominantly with muon-flavored neutrinos, previously unexplored parameter space can be excluded for masses between 30 and 150 MeV. Additionally, we make our Monte Carlo tools publicly available.
\end{abstract}

\maketitle

%%%%%%%%%%%%%%%%%%%%%%%%%%%%%%%%%%%%%%%%%
\textbf{Introduction --} The revelation that neutrinos are massive particles has spurred great interest in their properties. Perhaps the simplest and most elegant mass mechanisms for neutrinos are based on the seesaw concept~\cite{Mohapatra:1986bd, GellMann:1980vs, Yanagida:1979as, Mohapatra:1979ia, Schechter:1980gr, Magg:1980ut, Mohapatra:1979ia, Mohapatra:1980yp, Lazarides:1980nt, Ma:1998dx, Foot:1988aq, Ma:1998dn}, in which the scale of neutrino mass is suppressed with respect to the electroweak scale due to the presence of new particles and the breaking of lepton number.
In particular, several seesaw models postulate the existence of new leptons, neutral under the standard model gauge group --the heavy neutral leptons (HNLs).

\begin{figure}[b]
\centering
\includegraphics[width=\columnwidth]{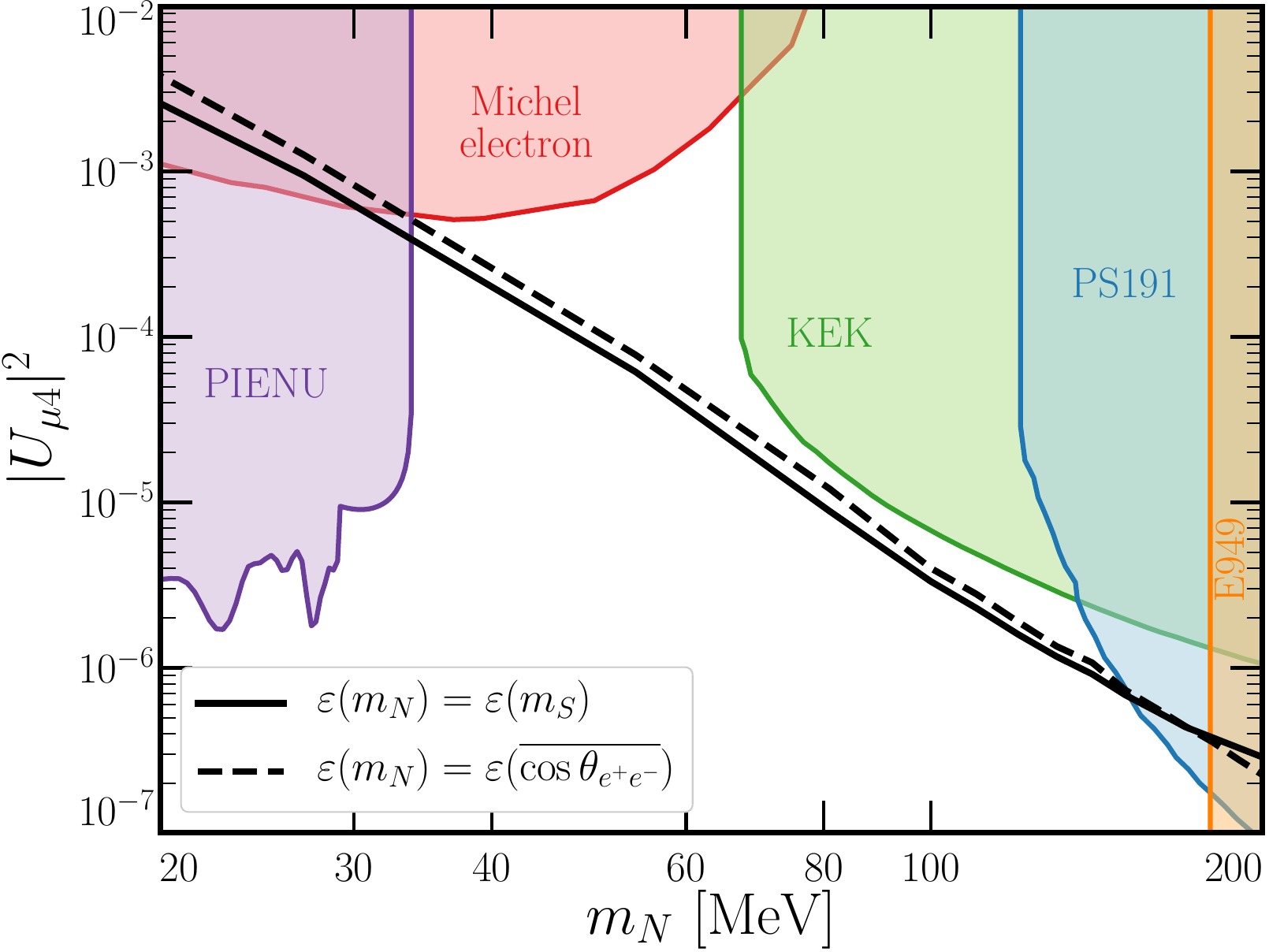}
\caption{MicroBooNE constraint on heavy neutral lepton parameter space as a function of its mass and mixing with muon neutrinos. Two assumptions regarding signal efficiency are shown in black (see text for detail) as well as comparisons against existing constraints~\cite{Aguilar-Arevalo:2019owf,Bryman:2019bjg,deGouvea:2015euy,Hayano:1982wu,Bernardi:1987ek,Artamonov:2014urb} (colored regions as labelled).\label{fig:Sensitivity}}
\end{figure}

In this letter, we show that current data from the MicroBooNE experiment~\cite{Abratenko:2021ebm} can constrain the existence of HNLs in an unexplored region of parameter space for masses between ${\sim}20-200$~MeV. In general, HNLs designate any fermion, neutral under the standard model gauge group, that mixes with neutrinos.
%More concretely, if HNLs are present, active neutrinos of well defined flavor could comprise an admixture of such heavy state,
%\begin{equation}
%  |\nu_\alpha \rangle = \sum_{i=1}^{3} U_{\alpha i}^*|\nu_i \rangle + \sum_K U_{\alpha K}^*|N_K \rangle,
%\end{equation}
%where $U$ is a generalized version of the usual Pontecorvo-Maki-Nakagawa-Sakata (PMNS) matrix, $N_K$ are the HNLs and we have explicitly split the sum in mostly active neutrinos $\nu_i$ and heavy states.
Although HNLs are a common, generic outcome of many seesaw models, there is no clear indication of what are their masses, neither how large are their mixings with active neutrino flavors.
This unpredictability has inspired a large effort to search for HNLs in a variety of environments:  kinematics of meson decays~\cite{Shrock:1981wq,Bernardi:1987ek, Aguilar-Arevalo:2015cdf, Aguilar-Arevalo:2017vlf, Sadovsky:2017qsr, Lazzeroni:2017fza, CortinaGil:2017mqf, Aguilar-Arevalo:2019owf, NA62:2020mcv, CortinaGil:2021gga}; production and decay in beam dump and neutrino experiments~\cite{Gorbunov:2007ak, Artamonov:2014urb, Alekhin:2015byh, Ballett:2016opr, Ballett:2019bgd, Abe:2019kgx, Berryman:2019dme,deGouvea:2021ual}; beta-decay spectral distortions~\cite{Schreckenbach:1983cg, Deutsch:1990ut, Hiddemann:1995ce, Derbin:1997ut, Holzschuh:1999vy, Holzschuh:2000nj, Galeazzi:2001py, Belesev:2013cba}; direct production at high energy colliders~\cite{Bergsma:1985is, CooperSarkar:1985nh, Abreu:1996pa, Aaboud:2019wfg, Sirunyan:2018xiv, Aaij:2020ovh,Blondel:2021mss}; and so on (see, e.g., Refs~\cite{Atre:2009rg, deGouvea:2015euy, Bryman:2019bjg, Bolton:2019pcu} for compilations of constraints).
In this work, we are interested in recasting the recent MicroBooNE Higgs-Portal Scalar (HPS) analysis~\cite{Abratenko:2021ebm} to constrain HNLs in the 20-200~MeV mass range, as exhibited by Fig.~\ref{fig:Sensitivity}. We include \href{\codeloc}{a github repository} with simulation code for generating signal events for both HNL and HPS decays in this setup~\cite{GitHubCode}.

Before going into details of our recasting, it is useful to describe the aforementioned search by the MicroBooNE experiment, a liquid argon time projection chamber detector in the Booster Neutrino Beam line~\cite{Antonello:2015lea}.
In Higgs portal models~\cite{Patt:2006fw, Batell:2019nwo, Egana-Ugrinovic:2019wzj, Dev:2019hho, Archer-Smith:2020hqq, Foroughi-Abari:2020gju}, a new, light scalar $S$ mixes with the Higgs, acquiring small couplings to all standard model fermions.
This mixing leads to production of the light scalars via loop effects in meson decays such as $K^+\to\pi^+S$, as well as decay of the light scalar into standard model fermions.
The MicroBooNE analysis searched for $S\to e^+e^-$ events which would be the dominant signature of the decay of this light scalar if its mass is below the dimuon production threshold.

The main experimental difficulty here lies in the fact that neutrino interactions  in argon that lead to electromagnetic showers, such as $\pi^0$ and photon production or $\nu_e$ charged current interactions, are a background to this search.
In view of that, MicroBooNE uses a clever, unorthodox strategy: to look for events in phase with the NuMI beam which points $8^\circ$ away from MicroBooNE.
In NuMI, charged kaons are produced when protons hit a target, and then travel through a decay pipe.
Most $K^+$ that do not decay will reach the absorber, stop, and decay at rest.
The absorber is located 100~meters from the MicroBooNE detector and at an angle of about $125^\circ$ with respect to the Booster Neutrino Beam line.
The $8^\circ$ off-axis angle between the NuMI beam and MicroBooNE reduces the neutrino flux from NuMI at MicroBooNE. 
Moreover, the large angle between the NuMI absorber and the detector relative to the NuMI beam direction  can be used to further mitigate backgrounds. 
These two features allows for a competitive experimental sensitivity to Higgs Portal Scalars (HPS).

MicroBooNE exploits several kinematical features of the $e^+e^-$ pair to increase the signal-to-background ratio. 
These include the opening angle between the electrons and their angles with respect to the line connecting the absorber and the detector center\footnote{The MicroBooNE detector does not distinguish electrons from positrons, so we will use ``electron'' whenever we refer to the $e^+$ or $e^-$ experimental signature.}, and the number of hits and length of the reconstructed objects in the LArTPC. 
No attempt to perform particle identification or energy reconstruction is made.

The key to our recast lies in the fact that HNLs would be produced in the NuMI absorber and detected at MicroBooNE by processes very similar to those of the HPS scenario.
Stopped kaons could decay isotropically to HNLs via $\nu_\mu$ mixing, namely $K^+\to\mu^+ N$.
Then, these HNLs could reach MicroBooNE and decay inside the detector.
In particular, the neutral current decay $N\to\nu_\mu e^+e^-$, via the same mixing, would lead to $e^+e^-$ pairs in MicroBooNE which would have similar kinematic features to the HPS scenario.
The very same data used to constrain HPS can be readily used to put a competitive constraint on as-yet unexplored values of HNL masses and mixings.

\textbf{Heavy neutral lepton model --} As discussed, there is no clear-cut indication of what would be the HNL masses and mixings, or even how many of them could exist in nature.
In light of that, we take a typical, simplified model approach which has the advantage of being general.
We consider a single HNL which mixes significantly only with muon neutrinos, that is
\begin{equation}
  |\nu_\mu \rangle = \sum_{i=1}^{3} U_{\mu i}^*|\nu_i \rangle + U_{\mu 4}^*|N \rangle,
\end{equation}
where $U$ is a generalized version of the usual Pontecorvo-Maki-Nakagawa-Sakata matrix and $N$ denotes the HNL.
Nonzero $U_{\tau 4}$ would not change the phenomenology we are interested in.
Although mixing with electron neutrino could lead to the same experimental signatures we are focusing on, existing constraints from $\pi^\pm$ decays (for $m_N < m_{\pi^\pm} - m_e$) are stronger than those on $U_{\mu 4}$ for similar masses, which would make the MicroBooNE constraint on $U_{e4}$ not competitive in this analysis, as checked numerically.

A nonzero mixing with muon neutrinos induces kaon branching ratio to $N$
\begin{equation}
  \text{Br}(K\!\!\to\!\mu N) \simeq \text{Br}(K\!\!\to\!\mu \nu)|U_{\mu4}|^2\rho_N\!\!\left(\frac{m_\mu^2}{m_K^2},\,\frac{m_N^2}{m_K^2}\right),
\end{equation}
where we have approximated $|U_{\mu 4}|^2\ll1$, which is justified by existing experimental bounds; $m_\mu$, $m_K$ and $m_N$ are the masses of the muon, the kaon and the HNL; and~\cite{Shrock:1980vy}
\begin{equation}
  \rho_N(x,y) = \frac{[x+y-(x-y)]^2\sqrt{1+(x-y)^2-2(x+y)}}{x(1-x)^2}\nonumber
\end{equation}
is a phase space factor such that in the limit $m_{\mu,N}\ll m_K$ we have $\rho_N\to1$.
The mixing with muon neutrinos also controls the $N$ decay rate to $e^+e^-$.
This decay depends on the nature of the HNL (Majorana versus Dirac).
Assuming for concreteness that $N$ is a Majorana fermion, its partial decay width to $e^+e^-$ pairs is given by~\cite{Gorbunov:2007ak}
\begin{align}
  \Gamma(N\to\nu_\mu e^+e^-)&=2\frac{G_F^2 |U_{\mu4}|^2 m_N^5}{768\pi^3} \left(1 - 4 s_W^2 + 8 s_W^4\right) \nonumber \\ &\hspace{2.7cm}+ \mathcal{O}(m_e^2/m_N^2),
\end{align}
where terms proportional to the ratio of the electron to HNL mass are neglected and $s_W$ is the sine of the weak mixing angle.
We have also kept explicit the factor of 2 due to the Majorana nature of $N$.
This decay, for $20<m_N<200$~MeV corresponds to a ${\sim}10\%$ branching ratio of all $N$ decays. The other $90\%$ is mostly $N\to\nu\nu\nu$, which is unobservable.

%Since neutrino interactions in argon is a background to this search, MicroBooNE focus on $e^+e^-$ events in time with the NuMI beam.
%
%
% analysis on searching for light scalars mixed with the Higgs
 
%Recently, MicroBooNE has performed a search for Higgs

%%%%%%%%%%%%%%%%%%%%%%%%%%%%%%%%%%%%%%%%%

%%%%%%%%%%%%%%%%%%%%%%%%%%%%%%%%%%%%%%%%%
\textbf{Signal Production and Constraint Recasting --} The event rate $R_X$ of a new-physics particle $X$, produced in a two-body kaon decay-at-rest, traveling to MicroBooNE from the NuMI absorber, and decaying within, can be expressed as the product of the production flux times the probability of decay:
\begin{align}
R_{X} &= \Phi_{X} A \,P(X\!\to e^+\! e^-) \varepsilon(m_X), \label{eq:RateProd2}\\
\Phi_{X} &= \frac{N_{K\mathrm{DAR}} \mathrm{Br}\left(K^\pm \to X\right)}{4\pi D^2}, \\
P(X\!\to e^+\! e^-) &= \frac{1}{\gamma c\tau_X} \int_{D}^{D + L} e^{-\frac{z}{\gamma c \tau_{X}}} dz  \nonumber \\
&\quad\times\mathrm{Br}(X\!\to e^+\! e^-)
\end{align}
Here, $\Phi_X$ is the flux of $X$ at the MicroBooNE detector, assumed to be nearly constant over the extent of the detector; $A$ is the cross-sectional area as viewed by an incoming $X$ from the NuMI absorber; $P(X\!\to e^+\! e^-)$ is the probability of $X$ decaying within the detector volume to, in our case, $e^+e^-+\,$anything; and $\varepsilon(m_X)$ is the signal reconstruction efficiency, which may depend on the mass of the decaying particle. $N_{K\mathrm{DAR}}$ is the number of kaons decaying at rest in the NuMI absorber during the MicroBooNE data collection, $D$ is the absorber/detector distance (${\sim}100$ m), $L$ is the extent of the detector for an incoming $X$, $\gamma = E_X/m_X$ can be determined from the two-body production of $X$, and $\tau_{X}$ is the proper lifetime of the particle $X$.

If $X$ is long-lived ($c\tau_X \gg D,~L$) relative to the other relevant distance scales\footnote{This limit is well-satisfied for $\left\lvert U_{\mu 4}\right\rvert^2 < (m_N/100\ \mathrm{MeV})^{-5}$. For the HPS, the requirement is $\sin^2\theta_S < 10^{-5}\ (m_S/100\ \mathrm{MeV})^{-1}$, satisfied for the parameter space of interest in Ref.~\cite{Abratenko:2021ebm}},
\begin{equation}
P(X\!\to e^+\! e^-) \simeq \frac{L}{\gamma c\tau_X}\mathrm{Br}\left(X\!\to e^+\! e^-\right) = \frac{L}{\gamma}\Gamma(X\!\to e^+\! e^-).
\end{equation}
When multiplied with $A$ in Eq.~\eqref{eq:RateProd2}, only the total volume of the detector, not its specific shape, enters the rate. We can compare the relative rates of hypothetical HPS and HNL decays within MicroBooNE coming from the NuMI absorber by calculating $R_N/R_S$:
\begin{align}
&\frac{R_{N}}{R_{S}} = \frac{\Phi_N}{\Phi_S} \frac{P(N\to \nu e^+ e^-)}{P(S \to e^+ e^-)} \frac{\varepsilon(m_N)}{\varepsilon(m_S)}, \label{eq:RelativeRate}\\
&= \frac{\mathrm{Br}\left(K^\pm \to \mu^\pm N\right)}{\mathrm{Br}\left(K^\pm \to \pi^\pm S\right)} \frac{m_N E_S \Gamma\left(N\to\nu e^+ e^-\right)}{m_S E_N \Gamma\left(S\to e^+ e^-\right)}\frac{\varepsilon(m_N)}{\varepsilon(m_S)}, \nonumber
\end{align}
where $E_{N,S}$ are the energies of the HNL and HPS.
This relative rate is described by the fluxes, probabilities of decaying within the detector, and signal efficiencies of the respective models. We defer discussion of the relative efficiencies to the next section.

\begin{figure}
\centering
\includegraphics[width=\columnwidth]{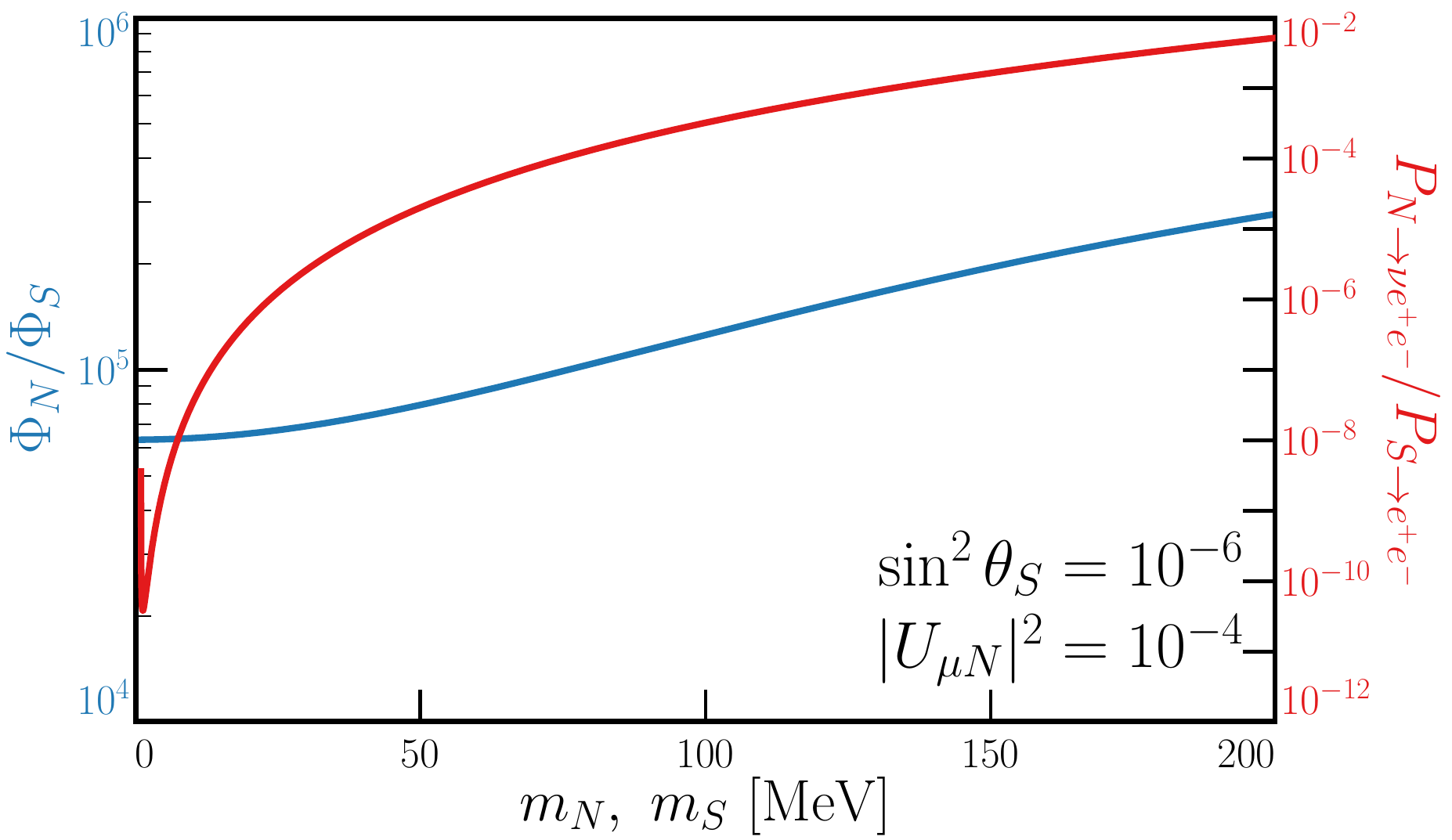}
\caption{Relative flux (blue) and decay probabilities (red) between the heavy neutral lepton $N$ and the Higgs portal scalar $S$ coming from the NuMI absorber via two-body kaon decay-at-rest and decaying within MicroBooNE. The product of these two, in addition to relative signal efficiencies, provides the relative signal event rate in MicroBooNE.\label{fig:FluxDecayRatio}}
\end{figure}

The first two fractions in Eq.~\eqref{eq:RelativeRate} can be determined from the HNL and HPS model parameters --- each is proportional to $\left\lvert U_{\mu N}\right\rvert^2/\sin^2\theta_S$. In Fig.~\ref{fig:FluxDecayRatio}, we present these two fractions (the relative fluxes in blue and the relative decay probabilities in red) as a function of the HNL or HPS mass, with benchmark values of $\left\lvert U_{\mu N}\right\rvert^2 = 10^{-4}$ and $\sin^2\theta_S = 10^{-6}$.
We see that, for these benchmark mixing values, a much larger flux of $N$ is produced relative to $S$, however, $S$ decays\footnote{The sharp mass-dependence of the red curve in Fig.~\ref{fig:FluxDecayRatio} comes from the relative scaling with masses of $\Gamma(N\to\nu e^+ e^-) \propto m_N^5$ vs. $\Gamma(S\to e^+ e^-) \propto m_S$.} much more rapidly than $N$ (effectively, $c\tau_{S} \gg c\tau_{N} \gg D$), leading to a much smaller probability that a given $N$ decays into an electron/positron pair in MicroBooNE than a given $S$ does. This combination implies that the HNL and HPS event rates can be comparable if one assumes similar signal reconstruction efficiencies. Ref.~\cite{Abratenko:2021ebm} provides constraints on $\theta_S$ and the efficiency $\varepsilon(m_S)$ for $m_S$ between $2m_e$ and $2m_\mu$: with reasonable assumptions about $\varepsilon(m_N)$, we can recast MicroBooNE's HPS constraint onto HNL parameter space.

%%%%%%%%%%%%%%%%%%%%%%%%%%%%%%%%%%%%%%%%%
\textbf{Event Kinematics \& Signal Efficiency --} The key challenge in recasting MicroBooNE's HPS search onto HNL parameter space is in determining the signal efficiency $\varepsilon(m_N)$. We take two approaches: first, we assume that $\varepsilon(m_N) = \varepsilon(m_S)$ and that events from an $N$ of a particular mass are accepted at the same rate as those from $S$ with the same mass. There are obvious shortcomings with this approach, however: because $N$'s decay is three-body, the kinematics of the $e^+e^-$ pair are different than those from the two-body $S \to e^+e^-$ decay. In Ref.~\cite{Abratenko:2021ebm}, signal identification (and background reduction) is optimized using boosted decision trees with kinematic quantities, especially the opening angle of the $e^+e^-$ pair and their angles relative to the direction from the NuMI absorber: no particle ID or calorimetry is used. 
Therefore, it seems more reasonable that the efficiency depends more strongly on the opening angle of the $e^+e^-$ pair than on the masses of the decaying particles.

To obtain a more realistic efficiency for HNL, we adopt the following second approach.
For each HPS mass, we map its efficiency onto the average cosine of the $e^+e^-$ opening angle $\overline{\cos\theta_{e^+e^-}}$. 
Then we assign the HNL efficiency to match the HPS efficiency for the same value of $\overline{\cos\theta_{e^+e^-}}$. 
Simulation code for three-body HNL decay $N\to \nu e^+ e^-$, including alternate model assumptions such as $N$ being a Dirac fermion, is available at \href{\codeloc}{this URL}~\cite{GitHubCode}.

\begin{figure}[b]
\centering
\includegraphics[width=0.95\columnwidth]{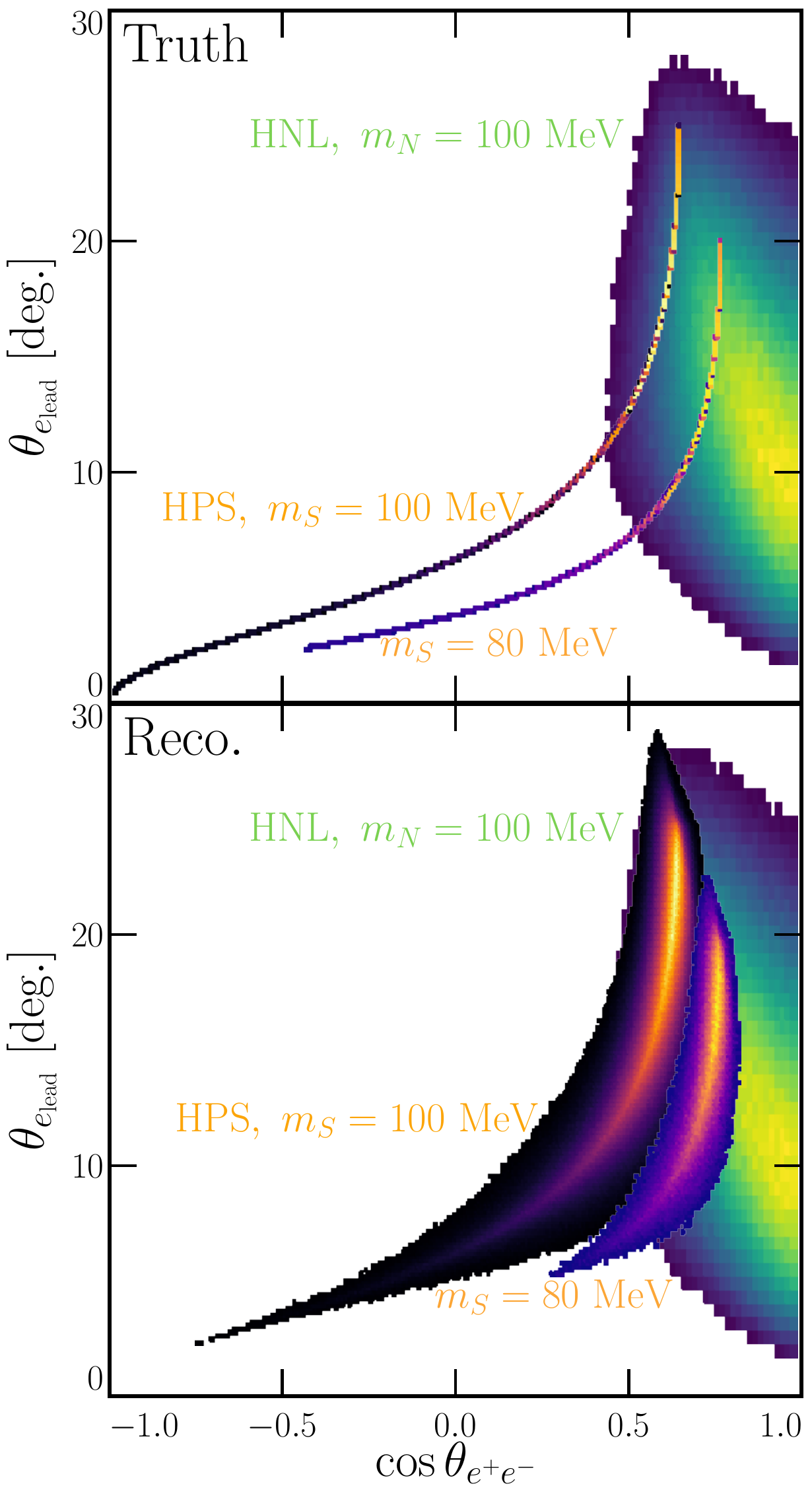}
\caption{Distributions of $e^+e^-$ pairs as a function of the cosine of their opening angle vs. the angle of the higher-energy particle relative to the NuMI absorber direction. Top: truth-level information, Bottom: reconstructed information after applying a 3$\degree$ angular uncertainty on electron tracks. Blue/green points come from an HNL with $m_N = 100$ MeV where the two purple/orange ones come from HPS with $m_S = 100$ MeV (top) and $m_S = 80$ MeV (bottom). In all three, brighter/darker colors correspond to more signal events.\label{fig:kinematics}}
\end{figure}

To justify this approach, Fig.~\ref{fig:kinematics}(top) displays several truth-level signal distributions as a function of the opening angle between the electron/positron pair $\cos\theta_{e^+e^-}$ and the angle between the higher-energy electron and the absorber direction $\theta_{e_{\rm lead}}$.
Three distributions are shown: the two purple/orange ones correspond to HPS with $m_S = 100$ and $80$ MeV, as labelled, and the blue/green one corresponds to an HNL with $m_N = 100$ MeV. 
For the HPS ones, we note that there is only one relevant kinematical quantity in the two-body $S$ decay, the rest-frame angle of the decay. 
This produces a strong correlation between these two quantities in the lab frame.
In all three distributions, brighter colors correspond to where more events are expected. 

For comparison, the bottom panel of Fig.~\ref{fig:kinematics} displays the same information after applying a $3\degree$ angular uncertainty to each of the electromagnetic showers~\cite{Adams:2019law}.
While we see a considerable overlap for $m_N = m_S$, there is even more overlap between $m_S=80$~MeV and $m_N=100$~MeV, as three-body decays prefer larger $\cos\theta_{e^+e^-}$ for the same mass of the mother particle.
Note that $m_S=80$~MeV and $m_N=100$~MeV exhibit approximately the same value of $\overline{\cos\theta_{e^+e^-}}$.
This overlap is especially obvious in the bottom panel in terms of the reconstructed angles.

\begin{figure}
\centering
\includegraphics[width=0.9\columnwidth]{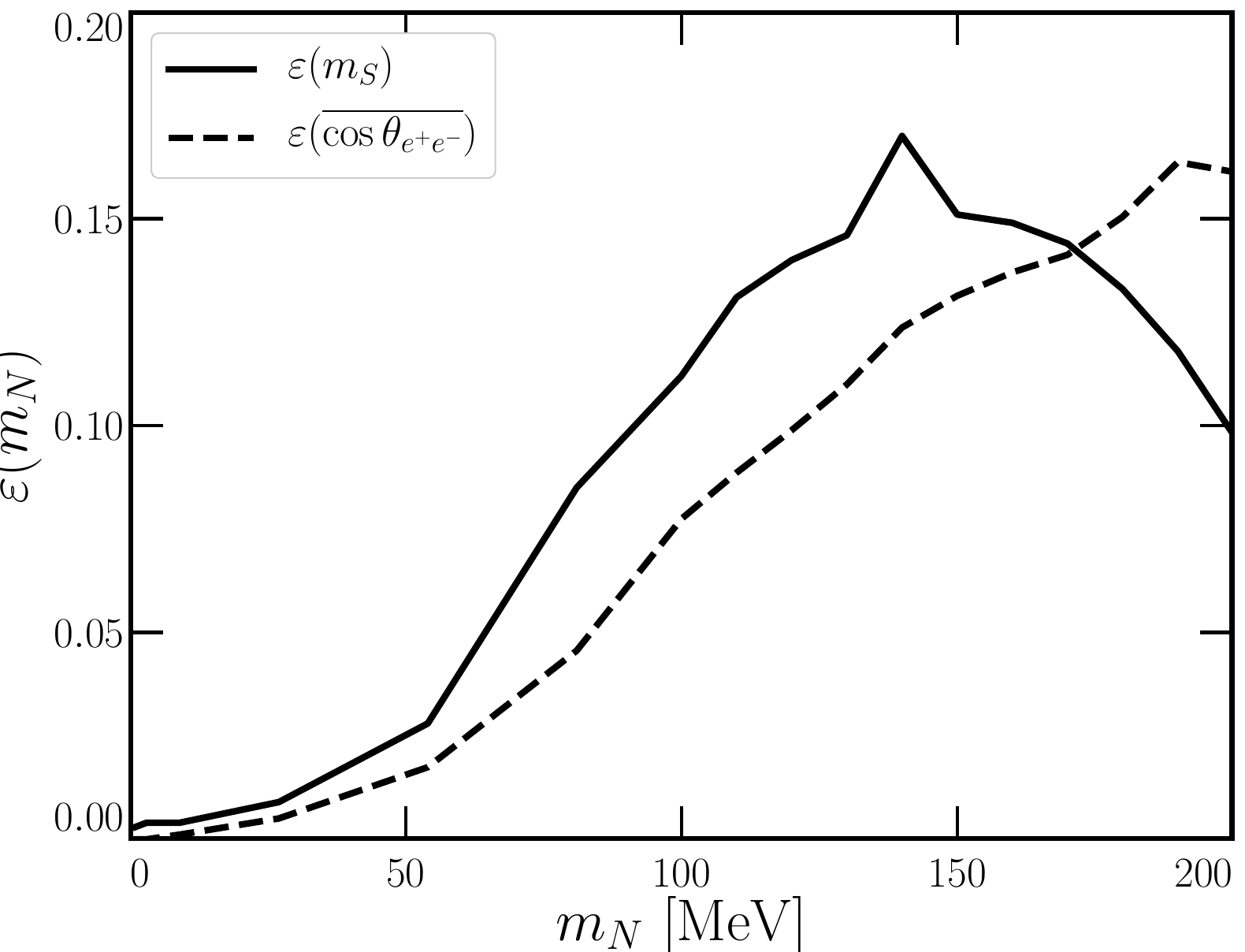}
\caption{Our two assumed signal efficiencies for HNL reconstruction: that the HNL efficiency matches the HPS efficiency given in Ref.~\cite{Abratenko:2021ebm} (solid) and that the dependence of the efficiency is based on the average opening angle of the electron/positron pair (dashed). \label{fig:efficiencies}}
\end{figure}

Finally, since HNLs decay to three daughter particles, the decay products tend to be less energetic. 
As the reconstruction of low energy electrons may be challenging, we also apply two kinematical cuts on top of both HPS and HNL efficiencies: $E_{e^\pm} > 10$ MeV, so that the two tracks can be reconstructed; and $\left\lvert \cos\theta_{e^+e^-}\right\rvert < \cos{(10\degree)}$, so that two distinct tracks can be identified.
%Although we may be double counting the detection threshold penalty, as it was already taking place in the HPS efficiency, this more conservative approach yields a very similar constrain on HNLs compared to the assumption $\varepsilon(m_S)=\varepsilon(m_N)$. \KJK{When calculating the effect of this cut, I also applied a similar cut to the HPS sample to account for that, so I don't think there's any ``double-counting'' here.}
Fig.~\ref{fig:efficiencies} shows the HNL efficiency as a function of the HNL mass for the two benchmark assumptions: $\varepsilon(m_S)=\varepsilon(m_N)$ (solid line) and $\varepsilon(m_N)$ tracing the average cosine of the $e^+e^-$ opening angle (dashed line).

\textbf{Results and discussion --} Under these two assumptions for $\varepsilon(m_N)$, in tandem with the efficiencies and limits on $\theta_S$ provided in Ref.~\cite{Abratenko:2021ebm}, we can use Eq.~\eqref{eq:RelativeRate} to determine the parameter space of $m_N$ vs. $\left\lvert U_{\mu4}\right\rvert^2$ for which a comparable signal rate is expected and therefore, MicroBooNE can set a constraint. This is shown in Fig.~\ref{fig:Sensitivity} with the solid line corresponding to $\varepsilon(m_N) = \varepsilon(m_S)$ and the dashed one to $\varepsilon(m_N) = \varepsilon(\overline{\cos\theta_{e^+e^-}})$. Colored regions depict existing experimental constraints as labelled from Refs.~\cite{Aguilar-Arevalo:2019owf,Bryman:2019bjg,deGouvea:2015euy,Hayano:1982wu,Bernardi:1987ek,Artamonov:2014urb}.\footnote{An additional constraint exists from PS191, for $30$ MeV $\lesssim m_N \lesssim 250$ MeV~\cite{Bernardi:1987ek}. However, as emphasized in Refs.~\cite{Kusenko:2004qc,Ruchayskiy:2011aa}, incorrect model assumptions were made in Ref.~\cite{Bernardi:1987ek}, requiring one to recast the presented limits appropriately. Given the lack of consistency in such recasts in the literature, possibly due to the lack of clarity in Ref.~\cite{Bernardi:1987ek} regarding systematic uncertainties and background events, we choose not to convey a constraint from this search in Fig.~\ref{fig:Sensitivity}. We use this confusion as further encouragement for MicroBooNE to explore this parameter space.} We see that MicroBooNE can constrain new parameter space for $34$ MeV $< m_N <  150$ MeV, including by over two orders of magnitude for $m_N \approx 65$ MeV. 
Even under the more conservative assumption on the HNL signal efficiency, the MicroBooNE constraint using current data remains world-leading in the aforementioned HNL mass window.

%%%%%%%%%%%%%%%%%%%%%%%%%%%%%%%%%%%%%%%%%
\textbf{Conclusions --} 
We have recasted MicroBooNE's search for Higgs portal scalars onto the heavy neutral lepton parameter space.
We have shown that, even with conservative assumptions on the HNL signal efficiency, MicroBooNE data still sets the world leading constraints on HNLs with masses between 34 and 150~MeV.
We encourage the MicroBooNE collaboration to use our publicly available \href{\codeloc}{Monte Carlo simulation code} to determine $\varepsilon(m_N)$ more exactly and carry out this analysis.
Finally, we call attention that a very similar search could be performed at ICARUS.
While its distance to the NuMI absorber is also about 100~meters, its volume is about 5 times larger than that of MicroBooNE.

%%%%%%%%%%%%%%%%%%%%%%%%%%%%%%%%%%%%%%%%%
%\section*{acknowledgments}
%%%%%%%%%%%%%%%%%%%%%%%%%%%%%%%%%%%%%%%%%

\begin{acknowledgments}
\textbf{Acknowledgements:} We thank Patrick Fox for valuable conversations on this work and Nikita Blinov and Shirley Li for testing our simulation code. We also thank Ryan Plestid and Manuel Gonz{\'a}lez L{\'o}pez for useful discussions regarding Ref.~\cite{Bernardi:1987ek}. Fermilab is operated by the Fermi Research Alliance, LLC under contract No. DE-AC02-07CH11359 with the United States Department of Energy. This project has received support from the European Union's Horizon 2020 research and innovation program under the Marie Sk\l{}odowska-Curie grant agreement No. 860881-HIDDeN.
\end{acknowledgments}

\bibliography{references}

\end{document}